\documentclass[aps,pra,preprint,groupedaddress]{revtex4-1}
\usepackage{amsmath}
\usepackage{graphicx}

\begin{document}


\title{Spinor wave equation, relativistic condition, and nonlocality of photon spin}


\author{Chun-Fang Li}
\email[]{cfli@shu.edu.cn}
\affiliation{Department of Physics, Shanghai University, 99 Shangda Road, 200444 Shanghai, China}


\date{\today}

\begin{abstract}

The purpose of this paper is to derive the photon spin and to deduce its properties from a pair of quantum equations for the photon. To this end, Darwin's equations are reinterpreted so as to meet the need of the quantum mechanics of the photon. It is found that the photon wavefunction transforms under Lorentz transformation as a spinor. The relativistic nature of the photon is expressed through a constraint equation on the wavefunction in such a way that the wave equation, which takes on the form of the Schr\"{o}dinger equation, is not Lorentz covariant unless the constraint equation is taken into account. The wave equation predicts the existence of a kind of spin, an intrinsic degree of freedom. But the constraint equation makes the spin nonlocal in the sense that no unique local density exists for the spin in position space. The nonlocality of the photon spin is a reflection of the nonlocality of the photon itself.

\end{abstract}



\maketitle


\newpage

\section{Introduction}

Spin is one of the most important physical phenomena in nature. Both electrons and photons are endowed with spin. The existence of the electron spin is predicted by Dirac's relativistic quantum equation~\cite{Dira}. But up till now, there has not been a generally-accepted relativistic quantum equation that predicts the existence of the photon spin.
The spin of the photon was theoretically interpreted  through separating its total angular momentum into spin and orbital parts~\cite{Darw, van-N1, van-N2, Li09, Li16E, Blio-AOA, Barn10, Bial-B11}. In the early days, such a separation was considered to be physically meaningless~\cite{Akhi-B, Cohen-DG, Barn-A, Barn02} due to the transversality conditions,
\begin{equation}\label{ME1}
    \nabla \cdot \boldsymbol{\mathcal E}=0, \quad \nabla \cdot \boldsymbol{\mathcal H}=0,
\end{equation}
on the electric and magnetic fields, $\boldsymbol{\mathcal E}$ and $\boldsymbol{\mathcal H}$, of free radiations.
Nowadays, much attention~\cite{Came, Blio-DN, Blio-BN14, Blio-N15, Bial, Barn14, Barn-AC, Lead, Lead18, Li-Z} was paid to the physical reality of the local density for the separated spin in position space.
So is it possible to formulate a relativistic quantum equation that not only predicts the existence of the photon spin but also shows its physical properties? The purpose of this paper is to address this question.

As a matter of fact, considerable efforts~\cite{Kell, Ritc, Mohr} have been made in seeking a relativistic quantum equation for the photon since the advent of quantum mechanics. The difficulty to reach a consensus on the form of a generally-accepted equation is mainly ascribed~\cite{Cook, Inag, Bial96, Kell} to the nonlocality~\cite{Jauch-P, Amre, Rose-S} of the photon in position space. In particular, it was argued~\cite{Akhi-B, Kell, Pauli}, on the basis of the assumption that the position-representation wavefunction of a quantum particle should be the probability amplitude for its position, that the nonlocality of the photon made it impossible to introduce the notion of photon wavefunction in position representation.
It is noted, however, that the above-mentioned assumption is made on the case in which the wavefunction only needs to satisfy the wave equation, the Schr\"{o}dinger equation, free of any additional constraints.
But as is known, the electric and magnetic fields of a free radiation need to satisfy the transversality conditions (\ref{ME1}) as well as the coupled equations,
\begin{equation}\label{ME2}
    \varepsilon_0 \frac{\partial \boldsymbol{\mathcal E}}{\partial t}
   =\nabla \times \boldsymbol{\mathcal H},                                           \quad
    \mu_0 \frac{\partial \boldsymbol{\mathcal H}}{\partial t}
   =-\nabla \times \boldsymbol{\mathcal E}.
\end{equation}
By this it is meant that after quantization, the photon wavefunction may have to satisfy, apart from a wave equation, an additional constraint equation.
Indeed, as early as in 1932 Darwin~\cite{Darw} cast the free-space Maxwell equations (\ref{ME1})-(\ref{ME2}) into two equations about a six-component wavefunction $\Psi(\mathbf{x},t)$,
\begin{subequations}\label{REoM}
\begin{align}
  i \hbar \frac{\partial \Psi}{\partial t}  & =H \Psi,   \label{DLE} \\
  (\mathbf{\Gamma} \cdot \mathbf{p})^2 \Psi & =p^2 \Psi, \label{RQC}
\end{align}
\end{subequations}
referred to as Darwin's equations, where
$H=ic \Gamma_0 (\mathbf{\Gamma} \cdot \mathbf{p})$,
$c=1/(\varepsilon_0 \mu_0)^{1/2}$ is the speed of light in vacuum,
\begin{equation*}
    \Gamma_0=\bigg(
                   \begin{array}{cc}
                     I_3 & 0 \\
                     0   & -I_3
                   \end{array}
             \bigg),                \quad
    \mathbf{\Gamma}=\bigg(
                          \begin{array}{cc}
                            0               & \mathbf{\Sigma} \\
                            \mathbf{\Sigma} & 0
                          \end{array}
                    \bigg),
\end{equation*}
$I_3$ is the 3-by-3 unit matrix, $(\Sigma_k)_{ij} =-i \epsilon_{ijk}$ with $\epsilon_{ijk}$ the Levi-Civit\'{a} pseudotensor, and
$\mathbf{p}=-i \hbar \nabla$
is the momentum operator.
The matrices $\Gamma_0$ and $\mathbf{\Gamma}$ are all Hermitian, having the following properties,
\begin{subequations}
\begin{align}
  \Gamma^2_0 & = 1,                                           \label{p1} \\
  \Gamma_0 \mathbf{\Gamma} +\mathbf{\Gamma} \Gamma_0 & =0,    \label{p2} \\
  \Gamma_i \Gamma_j \Gamma_k +\Gamma_k \Gamma_j \Gamma_i &
 =\Gamma_i \delta_{jk} +\Gamma_k \delta_{ij}, \quad  i=1,2,3. \label{p3}
\end{align}
\end{subequations}
It is seen that apart from satisfying the time-dependent wave equation (\ref{DLE}), the wavefunction $\Psi$ is also constrained by the time-independent equation (\ref{RQC}). As Darwin remarked, this is one peculiar feature that does not usually occur in quantum mechanics.
The key point is, as will be clear, that only when the constraint equation (\ref{RQC}), which is individually Lorentz covariant, is taken into account can the wave equation (\ref{DLE}) be Lorentz covariant. That is to say, the constraint equation (\ref{RQC}) shows up as a condition for the wave equation (\ref{DLE}) to be relativistic, called the relativistic condition (RC).
It is the RC that underlies the nonlocality of the photon in such a way that the integral of the modulus squared of the wavefunction over the whole position space gives the total probability of the photon but the wavefunction itself does not mean the probability amplitude for the position of the photon.

Frankly speaking, with Darwin's wavefunction that consists simply of the electric and magnetic fields, Eqs. (\ref{REoM}) cannot be the quantum equations for the photon. They can only be viewed as a modified version~\cite{Barn14} of Maxwell's equations (\ref{ME1})-(\ref{ME2}).
But I will show that once the wavefunction is properly connected with the electric and magnetic fields, they will meet the need of the quantum mechanics of the photon.
More important is that in contrast with the field-strength tensor \cite{Jack}, the wavefunction does not transform under Lorentz tranformation as a tensor. Instead, it transforms as a spinor \cite{Barn14} similar to the electron wavefunction in Dirac's equation.
Wave equation (\ref{DLE}) for free photons is analogous to Dirac's equation for free electrons. It takes on the form of the Schr\"{o}dinger equation with $H$ the Hamiltonian.
The same as Dirac's equation predicts the existence of the spin of the electron, it predicts the existence of a kind of spin.
If the RC (\ref{RQC}) is not considered, the predicted spin appears to be an intrinsic degree of freedom and is represented by the operator
$-i \mathbf{\Gamma} \times \mathbf{\Gamma} $, denoted by $\mathbf \Omega$.
After the role of the RC is taken into consideration, the spin, which is now the spin of the photon, becomes non-intrinsic~\cite{Bial-B12}. Its representative operator changes into
$(\mathbf{\Omega} \cdot \mathbf{p}) \mathbf{p}/p^2 $.
It is not only oriented in the propagation direction~\cite{Jauch-R} but also has commuting components~\cite{van-N1, van-N2}.
Nevertheless, contrary to the claim by Bliokh \textit{et al}~\cite{Blio-DN, Blio-N15, Blio-BN14}, the occurrence of factor $ 1/p^2 $ in the operator conveys the nonlocality of the photon spin in the sense that there does not exist a unique local density for the photon spin in position space~\cite{Barn10, Bial, Li-Z}.
The nonlocality of the photon spin is a reflection of the nonlocality of the photon itself.
All these results make up the main content of this paper.
Let us first investigate what the RC (\ref{RQC}) exactly means to the Lorentz covariance of the wave equation (\ref{DLE}).

\section{Lorentz Covariance of wave equation and RC}\label{LCoDE}

For the sake of clarity, I start with the Lorentz covariance of Maxwell's equations (\ref{ME1})-(\ref{ME2}).
As is known, by denoting
$(ict,x,y,z) \equiv (x_0,x_1,x_2,x_3)$,
the first equation in (\ref{ME1}) and the first equation in (\ref{ME2}) can be combined together into
\begin{equation}\label{TE-D}
    \frac{\partial F_{\mu \nu}}{\partial x_\nu } =0, \quad \mu,\nu=0,1,2,3,
\end{equation}
in terms of the $4 \times 4$ antisymmetric matrix~\cite{Cook}
\begin{equation*}
F_{\mu \nu}=
\left(
      \begin{array}{cccc}
         0                 & i\mathcal{E}_1/c     & i\mathcal{E}_2/c     & i\mathcal{E}_3/c \\
        -i \mathcal{E}_1/c & 0                    & \mu_0 \mathcal{H}_3  & -\mu_0 \mathcal{H}_2 \\
        -i \mathcal{E}_2/c & -\mu_0 \mathcal{H}_3 & 0                    & \mu_0 \mathcal{H}_1 \\
        -i \mathcal{E}_3/c & \mu_0 \mathcal{H}_2  & -\mu_0 \mathcal{H}_1 & 0 \\
      \end{array}
\right),
\end{equation*}
where summation convention has been assumed. Similarly, the second equation in (\ref{ME1}) and the second equation in (\ref{ME2}) can be combined together into
\begin{equation}\label{TE-C}
    \frac{\partial F_{\mu \nu}}{\partial x_\lambda }
   +\frac{\partial F_{\lambda \mu}}{\partial x_\nu }
   +\frac{\partial F_{\nu \lambda}}{\partial x_\mu }  =0.
\end{equation}
Since the matrix $F_{\mu \nu}$ transforms under Lorentz transformation as a tensor of the second rank, known as the field-strength tensor, Eqs. (\ref{TE-D}) and (\ref{TE-C}) are all Lorentz covariant individually.

To be sure that the wave equation (\ref{DLE}) is Lorentz covariant, it is instructive to show that Darwin's equations (\ref{REoM}) can be cast into the form of Maxwell equations (\ref{ME1})-(\ref{ME2}).
Letting
$\Psi=\frac{1}{\sqrt 2} \bigg(
                              \begin{array}{c}
                                \mathbf{F}_u \\
                                \mathbf{F}_l
                              \end{array}
                        \bigg)$
in accordance with the concrete forms of the matrices $\Gamma_0$ and $\mathbf \Gamma$, where the factor $\frac{1}{\sqrt 2}$ is introduced for later convenience, one readily changes Eq. (\ref{DLE}) into
\begin{equation*}
    i \hbar \frac{\partial \mathbf{F}_u}{\partial t}=-c \mathbf{p} \times \mathbf{F}_l, \quad
    i \hbar \frac{\partial \mathbf{F}_l}{\partial t}= c \mathbf{p} \times \mathbf{F}_u,
\end{equation*}
or, equivalently,
\begin{equation}\label{CEoM}
    \frac{\partial \mathbf{F}_u}{\partial t}= c \nabla \times \mathbf{F}_l, \quad
    \frac{\partial \mathbf{F}_l}{\partial t}=-c \nabla \times \mathbf{F}_u,
\end{equation}
where the relation
$(\mathbf{\Sigma} \cdot \mathbf{a}) \mathbf{b}= i \mathbf{a} \times \mathbf{b}$
has been used. They are the same as Maxwell's equations (\ref{ME2}) if the following correspondences are assumed,
\begin{equation}\label{corr}
    \mathbf{F}_u \sim \sqrt{\varepsilon_0} \boldsymbol{\mathcal E}, \quad
    \mathbf{F}_l \sim \sqrt{\mu_0}         \boldsymbol{\mathcal H}.
\end{equation}
Meanwhile, Eq. (\ref{RQC}) can be rewritten in terms of $\mathbf{F}_u$ and $\mathbf{F}_l$ as
\begin{equation}\label{RCC}
    \mathbf{p} (\mathbf{p} \cdot \mathbf{F}_u)=0, \quad
    \mathbf{p} (\mathbf{p} \cdot \mathbf{F}_l)=0,
\end{equation}
or, equivalently,
\begin{equation}\label{RCC-r}
    \nabla (\nabla \cdot \mathbf{F}_u)=0, \quad
    \nabla (\nabla \cdot \mathbf{F}_l)=0.
\end{equation}
Since $H^2 =c^2 p^2$ by virtue of Eq. (\ref{RQC}), a photon with nonzero energy cannot have vanishing momentum. So when the energy does not vanish, Eqs. (\ref{RCC}) mean
$\mathbf{p} \cdot \mathbf{F}_u =0$ and $\mathbf{p} \cdot \mathbf{F}_l =0$.
That is to say,
\begin{equation}\label{EoD}
    \nabla \cdot \mathbf{F}_u =0, \quad \nabla \cdot \mathbf{F}_l =0.
\end{equation}
They are the same as Maxwell's equations (\ref{ME1}). Moreover, when the energy vanishes, in other words, when the wavefunction satisfies
$H \Psi =0$, one must have
$\frac{\partial \Psi}{\partial t} =0$
or, equivalently,
\begin{equation*}
    \nabla \times \mathbf{F}_u =0, \quad \nabla \times \mathbf{F}_l =0,
\end{equation*}
in accordance with Eqs. (\ref{CEoM}). With the help of these equations, one readily obtains from Eqs. (\ref{RCC-r})
\begin{equation}\label{LE}
    \nabla^2 \mathbf{F}_u =0, \quad \nabla^2 \mathbf{F}_l =0.
\end{equation}
According to Stratton~\cite{Stra}, a vector function that throughout all space satisfies Laplace's equation vanishes at infinity so that
$\mathbf{F}_u |_{\infty} =\mathbf{F}_l |_{\infty} =0$. Furthermore, from the uniqueness theorem~\cite{Jack} it follows that Eqs. (\ref{LE}) with these boundary conditions have only trivial solutions,
$\mathbf{F}_u =0$ and $\mathbf{F}_l =0$,
which satisfy Eqs. (\ref{EoD}).
Equation (\ref{RQC}) is thus cast into the form of Maxwell equations (\ref{ME1}) whether the energy vanishes or not.
It is also seen from the casting process that no photon can have vanishing energy or vanishing momentum.

Now that Darwin's equations (\ref{REoM}) can be cast into the form of Maxwell equations (\ref{ME1})-(\ref{ME2}), one might infer the Lorentz covariance of the wave equation (\ref{DLE})
from the Lorentz covariance of tensor equations (\ref{TE-D})-(\ref{TE-C}).
But unexpectedly, the wave equation is not Lorentz covariant by itself. Indeed, Eq. (\ref{DLE}) can be rewritten as
\begin{equation*}
    \Gamma_\mu p_\mu \Psi=0,
\end{equation*}
where
$p_0=-i \hbar \frac{\partial}{\partial x_0}$.
Multiplying this equation by $\Gamma_\nu p_\nu$ on the left and using Eqs. (\ref{p1}) and (\ref{p2}), one has
\begin{equation*}
    [p^2_0 +(\mathbf{\Gamma} \cdot \mathbf{p})^2] \Psi=0.
\end{equation*}
It is apparently not Lorentz covariant. However, upon substituting Eq. (\ref{RQC}), one gets the following Klein-Gordon equation for zero-mass particles,
\begin{equation*}
    p_\mu p_\mu \Psi=0.
\end{equation*}
That is to say, the wave equation is Lorentz covariant so long as Eq. (\ref{RQC}) is taken into account. This shows that Eq. (\ref{RQC}) acts as a condition for the wave equation to be relativistic.
In a word, the Lorentz covariance of wave equation (\ref{DLE}) does not follow directly from the Lorentz covariance of tensor equations (\ref{TE-D})-(\ref{TE-C}). The key point here is that in contrast with the field-strength tensor $F_{\mu \nu}$, the wavefunction does not transform under Lorentz transformation as a tensor.

To look in more detail at how the RC (\ref{RQC}) makes the wave equation (\ref{DLE}) relativistic, the transformation law for the wavefunction under Lorentz transformation is given in Appendix \ref{LT} and the Lorentz covariance of Darwin's equations (\ref{REoM}) is proven in Appendix \ref{LC}.
Equation (\ref{LT-WF}) shows that the wavefunction transforms as a spinor instead of as a tensor. This does not mean that it transforms in a nonlocal fashion as Cook~\cite{Cook} discussed. As a matter of fact, as is explicitly shown by Eqs. (\ref{LT-1}) and (\ref{LT-2}), the upper and lower parts of the wavefunction transform in the same way as the electric and magnetic fields transform.
Also noteworthy is, as will be discussed at the end of Section \ref{NP}, that for Darwin's equations to be the relativistic quantum equations for the photon, the upper and lower parts of the wavefunction do not locally depend on the electric and magnetic fields.

From Darwin's equations (\ref{REoM}) one can derive many results like those for the electron from Dirac's equation. The present paper is mainly concerned about how the photon spin and its physical properties are derived.
It will been seen in the next section that even though the wave equation (\ref{DLE}) is not relativistic by itself, it predicts the existence of a kind of spin the same as Dirac's equation predicts the existence of the electron spin.

\section{Spin Predicted Solely by wave Equation}\label{CanoS}

Putting the RC (\ref{RQC}) aside, it is not difficult to show by use of the wave equation (\ref{DLE}) that the orbital angular momentum
$\mathbf{L}=\mathbf{x} \times \mathbf{p}$
is not a constant of motion,
\begin{equation*}
    [H, \mathbf{L}]=\hbar c \Gamma_0 \mathbf{\Gamma} \times \mathbf{p},
\end{equation*}
where the following commutation relations are assumed,
\begin{equation}\label{CR-PandX}
    [p_i, p_j] =0, \quad [x_i, p_j] =i \hbar \delta_{ij}.
\end{equation}
For this reason, one introduces a constant vector matrix
$-i \mathbf{\Gamma} \times \mathbf{\Gamma}$, which is
$
\mathbf{\Omega}=\bigg(\begin{array}{cc}
                        \mathbf{\Sigma} & 0 \\
                        0 &               \mathbf{\Sigma}
                      \end{array}
                \bigg)
$
by virtue of the commutation relation
\begin{equation}\label{CR-Sigma}
    [\Sigma_i, \Sigma_j]=i \epsilon_{ijk} \Sigma_k.
\end{equation}
With the help of Eqs. (\ref{p2}) and (\ref{p3}), it is easy to find
\begin{equation*}
    [H, \mathbf{\Omega}]=-c \Gamma_0 \mathbf{\Gamma} \times \mathbf{p},
\end{equation*}
indicating that the sum of $\mathbf L$ and $\hbar \mathbf{\Omega}$ is a constant of motion.
Because it is independent of the extrinsic degrees of freedom such as the momentum, $\mathbf \Omega$ represents an intrinsic degree of freedom, called the spin. It obeys the canonical commutation relation,
\begin{equation}\label{CR-Omega}
    [\Omega_i, \Omega_j]=i \epsilon_{ijk} \Omega_k,
\end{equation}
by virtue of Eq. (\ref{CR-Sigma}). The constant of motion,
$\mathbf{L}+\hbar \mathbf{\Omega}$,
is the total angular momentum.
According to the definition~\cite{Saku}, the expectation value of the spin in an arbitrary state $\Psi$ that is normalized as
$\int \Psi^\dag \Psi d^3 x=1$
is given by
\begin{equation}\label{EoS}
    \langle \mathbf{\Omega} \rangle =\int \Psi^\dag \mathbf{\Omega} \Psi d^3 x
\end{equation}
in units of $\hbar$.

Being an intrinsic degree of freedom, the spin here can also be represented in momentum representation by the same constant vector operator $\mathbf \Omega$.
In fact, denoting by $\psi(\mathbf{k},t)$ the wavefunction in momentum representation with $\mathbf k$ the wavevector, which is the Fourier component of $\Psi$,
\begin{equation}\label{FT}
    \Psi(\mathbf{x}, t)=\frac{1}{(2 \pi)^{3/2}}
    \int \psi(\mathbf{k}, t) \exp(i \mathbf{k} \cdot \mathbf{x}) d^3 k,
\end{equation}
one readily changes Eq. (\ref{EoS}) into
\begin{equation}\label{EoS-k}
    \langle\mathbf{\Omega}\rangle=\int \psi^\dag \mathbf{\Omega} \psi d^3 k.
\end{equation}
In a word, wave equation (\ref{DLE}) predicts the existence of the spin $\mathbf \Omega$ as long as commutation relations (\ref{CR-PandX}) are satisfied.

Nevertheless, it is noted that the constant operator $\mathbf \Omega$ does not represent the spin of the photon.
This is because the canonical commutation relation (\ref{CR-Omega}) together with the property
$\mathbf{\Omega}^2=2$
leads to a consequence~\cite{Saku} that the component of $\mathbf \Omega$ in any fixed direction has eigenvalues of $\pm 1$ and $0$, which is apparently not the property of the photon spin.
Fortunately, the wavefunction of the photon has to obey, apart from the wave equation (\ref{DLE}), the RC (\ref{RQC}), which has not yet been exploited at all. Let us further examine how the RC determines the properties of the photon spin.

\section{Nonlocality of Photon Spin Determined by RC}

\subsection{RC in momentum representation}

Considering that the RC (\ref{RQC}) is expressed in terms of the momentum, it is beneficial to write it out in momentum representation.
To this end, one substitutes Eq. (\ref{FT}) into (\ref{REoM}) to get
\begin{subequations}\label{REoM-k}
\begin{align}
  i \hbar \frac{\partial \psi}{\partial t}  & =H   \psi,     \label{DLE-k} \\
  (\mathbf{\Gamma} \cdot \mathbf{k})^2 \psi & =k^2 \psi,     \label{RQC-k}
\end{align}
\end{subequations}
where
$H=i \hbar c \Gamma_0 (\mathbf{\Gamma} \cdot \mathbf{k})$
and $k=|\mathbf{k}|$.
Equation (\ref{RQC-k}) shows that $\psi$ is the eigenfunction of
$ (\mathbf{\Gamma} \cdot \mathbf{k})^2 $
with eigenvalue $k^2 $.
From this equation one has $H^2 =\hbar^2 c^2 k^2$, indicating that Eq. (\ref{DLE-k}) has solutions of negative as well as positive energy.
Akin to solutions to Dirac's equation~\cite{Pesk-S}, solutions of negative energy correspond to antiphotons~\cite{Kell} if solutions of positive energy correspond to photons.
In this paper, I am not concerned with photon annihilation and creation and hence will consider only solutions of positive energy. Taking this into account, Eq. (\ref{RQC-k}) is equivalent to the following eigenvalue equation of the Hamiltonian,
\begin{equation*}
    H \psi =\hbar \omega \psi,
\end{equation*}
where $\omega=c k$.
Letting
$
\psi=\frac{1}{\sqrt 2} \bigg(
                             \begin{array}{c}
                               \mathbf{f}_u \\
                               \mathbf{f}_l
                             \end{array}
                       \bigg)
$,
where $\mathbf{f}_u$ and $\mathbf{f}_l$ are the Fourier transformations of $\mathbf{F}_u$ and $\mathbf{F}_l$, respectively,
\begin{equation}\label{FT-f}
    \mathbf{f}_{u,l}(\mathbf{k},t)=\frac{1}{(2 \pi)^{3/2}}
    \int \mathbf{F}_{u,l}(\mathbf{x},t) \exp(-i \mathbf{k} \cdot \mathbf{x}) d^3 x,
\end{equation}
one readily changes the eigenvalue equation into
\begin{equation}\label{CE-k}
    \mathbf{w} \times \mathbf{f}_l= -\mathbf{f}_u , \quad
    \mathbf{w} \times \mathbf{f}_u=  \mathbf{f}_l ,
\end{equation}
where $\mathbf{w}=\mathbf{k}/k$ stands for the unit momentum. From these two equations it follows that
\begin{equation}\label{TC-k}
    \mathbf{w} \cdot \mathbf{f}_{u,l}=0,
\end{equation}
which is the same as the result of Eq. (\ref{RQC-k}) expressed in terms of $\mathbf{f}_u$ and $\mathbf{f}_l$.
In a word, the RC (\ref{RQC}) is converted into (\ref{CE-k}) or (\ref{TC-k}) in momentum representation.

\subsection{Photon spin operator in momentum representation}

Now it is ready to see how the RC affects the properties of the photon spin. Equation (\ref{TC-k}) tells that
\begin{equation*}
    \mathbf{f}^\dag_{u,l}(\mathbf{\Sigma} \times \mathbf{w})
    \mathbf{f}_{u,l}=0,
\end{equation*}
where the relation
$\mathbf{a}^\dag \mathbf{\Sigma} \mathbf{b} =-i \mathbf{a}^* \times \mathbf{b}$
\cite{Cohen-DG} has been used. Taking this equation into consideration and using the identity
\begin{equation*}
    \mathbf{\Sigma}=\mathbf{\Sigma} \cdot \mathbf{w} \mathbf{w}
               -(\mathbf{\Sigma} \times \mathbf{w}) \times \mathbf{w},
\end{equation*}
where $\mathbf{w} \mathbf{w}$ is a dyadic, one finds
\begin{equation}\label{redu-Si}
    \mathbf{f}^\dag_{u,l} \mathbf{\Sigma} \mathbf{f}_{u,l}
   =\mathbf{f}^\dag_{u,l} (\mathbf{\Sigma} \cdot \mathbf{w} \mathbf{w}) \mathbf{f}_{u,l} ,
\end{equation}
which is equivalent to
\begin{equation*}
    \psi^\dag \mathbf{\Omega} \psi
   =\psi^\dag (\mathbf{\Omega} \cdot \mathbf{w} \mathbf{w}) \psi.
\end{equation*}
Substituting it into Eq. (\ref{EoS-k}), one has
\begin{equation}\label{EoPS}
    \langle \mathbf{S} \rangle
   =\int\psi^\dag (\mathbf{\Omega} \cdot \mathbf{w} \mathbf{w}) \psi d^3 k.
\end{equation}
Based on the arbitrariness of $\psi$ one concludes that the RC (\ref{TC-k}) reduces the spin operator from $\mathbf \Omega$ to $\mathbf{\Omega} \cdot \mathbf{w} \mathbf{w}$.
This indicates that the operator for the photon spin in momentum representation is not the constant vector matrix $\mathbf \Omega$. Instead, it is
\begin{equation}
    \mathbf{S}=\mathbf{\Omega} \cdot \mathbf{w} \mathbf{w}.
\end{equation}
In the first place, it coincides with the well-known conclusion that the spin of the photon is always oriented in its propagation direction~\cite{Jauch-R}. In the second place, it has commuting Cartesian components,
\begin{equation*}
    [S_i, S_j]=0,
\end{equation*}
in perfect agreement with the result that was obtained in the framework of second quantization~\cite{van-N1, van-N2}.
In the third place, it shows that as a momentum-dependent quantity, the spin of the photon is not an independent degree of freedom~\cite{Bial-B12}.

It is worth noting that commuting with the Hamiltonian,
$[H, \mathbf{S}]=0$,
the photon spin is a constant of motion. More importantly, the occurrence of the dyadic $\mathbf{ww}$ in $\mathbf{S}$ denies the existence of the local density for the photon spin in position space as is shown below.

\subsection{Nonexistence of local density for photon spin in position space}

Substituting the inverse Fourier transformation of Eq. (\ref{FT}) into Eq. (\ref{EoPS}), one gets
\begin{equation*}
    \langle \mathbf{S} \rangle=\int \mathbf{s}(\mathbf{x},t) d^3 x,
\end{equation*}
where
\begin{equation}\label{s}
    \mathbf{s}(\mathbf{x},t)
   =\Psi^\dag (\mathbf{x},t) \int \mathbf{G} (\mathbf{x}-\mathbf{x}')
                                        \Psi (\mathbf{x}',t) d^3 x',
\end{equation}
\begin{equation*}
    \mathbf{G} (\mathbf{x}) =\frac{1}{(2 \pi)^3} \mathbf{\Omega} \cdot
       \int \mathbf{w} \mathbf{w} \exp(i \mathbf{k} \cdot \mathbf{x}) d^3 k.
\end{equation*}
The integrand $\mathbf s$ does not locally depend on the wavefunction $\Psi$. Its value at any particular point $\mathbf x$ depends not only on the value of the wavefunction at that point but also on the value at all other points.
It cannot, therefore, be interpreted as the local density for the photon spin in position space though its integral over the whole position space yields the expectation value.
It is observed that if $\mathbf{w}\mathbf{w}$ is replaced with the unit dyadic, $\mathbf{G} (\mathbf{x})$ will be replaced with
$\mathbf{\Omega} \delta^3 (\mathbf{x})$
and $\mathbf{s}(\mathbf{x}, t)$ will be replaced with
$\Psi^\dag \mathbf{\Omega} \Psi$. In that case, $\mathbf \Omega$ will represent the spin in position representation.
Since $\mathbf{w}\mathbf{w}$ is not the unit dyadic, $\mathbf \Omega$ cannot be the operator for the spin of the photon in position representation as Bliokh \textit{et al} claimed~\cite{Blio-DN, Blio-N15, Blio-BN14}.
The nonexistence of the local density for the photon spin can also be illustrated by showing that the expectation value of the photon spin is equal to the integral of different integrands over the whole position space.

Equation (\ref{EoPS}) can be rewritten in terms of the upper and lower parts of $\psi$ as
\begin{equation*}
    \langle \mathbf{S} \rangle =\frac{1}{2}
    \int[\mathbf{f}_u^\dag (\mathbf{\Sigma} \cdot \mathbf{w}) \mathbf{f}_u
        +\mathbf{f}_l^\dag (\mathbf{\Sigma} \cdot \mathbf{w}) \mathbf{f}_l] \mathbf{w} d^3 k.
\end{equation*}
Observing that
\begin{equation*}
    \mathbf{f}_u^\dag (\mathbf{\Sigma} \cdot \mathbf{w}) \mathbf{f}_u
   =\mathbf{f}_l^\dag (\mathbf{\Sigma} \cdot \mathbf{w}) \mathbf{f}_l
\end{equation*}
by virtue of Eq. (\ref{CE-k}), one has
\begin{equation*}
    \langle \mathbf{S} \rangle
   =\int [\mathbf{f}^\dag_{u,l} (\mathbf{\Sigma} \cdot \mathbf{w}) \mathbf{f}_{u,l}]
          \mathbf{w} d^3 k.
\end{equation*}
Resorting to the property (\ref{redu-Si}), one gets
\begin{equation}\label{EoS-f}
    \langle \mathbf{S} \rangle
   =\int \mathbf{f}^\dag_{u,l} \mathbf{\Sigma} \mathbf{f}_{u,l} d^3 k
   =-i \int \mathbf{f}^*_{u,l} \times \mathbf{f}_{u,l} d^3 k.
\end{equation}
Upon substituting the Fourier transformation (\ref{FT-f}), one finds
\begin{equation*}
    \langle \mathbf{S} \rangle
   =-i \int \mathbf{F}^*_{u,l} \times \mathbf{F}_{u,l} d^3 x,
\end{equation*}
which is expressed in terms of the position-representation wavefunction $\Psi$ as
\begin{equation*}
    \langle \mathbf{S} \rangle
   =\int \Psi^\dag (1 \pm \Gamma_0) \mathbf{\Omega} \Psi d^3 x.
\end{equation*}
From this expression one further deduces
\begin{equation*}
    \langle \mathbf{S} \rangle
   =\int \Psi^\dag \mathbf{\Omega} \Psi d^3 x,
\end{equation*}
which is the same as Eq. (\ref{EoS}). Generally speaking, neither
$\Psi^\dag (1 +\Gamma_0) \mathbf{\Omega} \Psi$
nor
$\Psi^\dag (1 -\Gamma_0) \mathbf{\Omega} \Psi$
is equal to
$\Psi^\dag \mathbf{\Omega} \Psi$
for any particular photon state $\Psi$.
It is thus concluded that there is no unique local density for the photon spin in position space.

It is pointed out, by the way, that due to the same RC (\ref{RQC}), the expectation value of the photon orbital angular momentum, defined by
\begin{equation}\label{EoL}
    \langle \mathbf{L} \rangle=\int \Psi^\dag \mathbf{L} \Psi d^3 x ,
\end{equation}
can be converted into
\begin{equation}\label{EoL-f}
    \langle \mathbf{L} \rangle
   =-i \hbar \int \mathbf{f}_{u,l}^\dag
                 (\mathbf{k} \times \nabla_{\mathbf k}) \mathbf{f}_{u,l} d^3 k
\end{equation}
in momentum representation or into
\begin{equation}\label{EoL-F}
    \langle \mathbf{L} \rangle
   =-i \hbar \int \mathbf{F}_{u,l}^\dag
                 (\mathbf{x} \times \nabla) \mathbf{F}_{u,l} d^3 x
\end{equation}
in position representation, where $\nabla_{\mathbf k}$ denotes the gradient operator with respect to $\mathbf k$.
Equation (\ref{EoL-F}) can be further expressed in terms of the wavefunction $\Psi$ as
\begin{equation*}
    \langle \mathbf{L} \rangle
   =\int \Psi^\dag (1 \pm \Gamma_0) \mathbf{L} \Psi d^3 x .
\end{equation*}
A comparison with Eq. (\ref{EoL}) shows that there does not exist such a notion as the local density for the photon orbital angular momentum in position space, indicating that the photon orbital angular momentum is also nonlocal in position space~\cite{Bial}.
It is thus seen that the transversality conditions (\ref{ME1}) on the electric and magnetic fields, which correspond to the RC (\ref{RQC}) on the wavefunction, do not mean the inseparability of the photon spin and orbital angular momentum. Instead, they mean the nonlocality of the photon spin and orbital angular momentum.
Such a nonlocality reflects the nonlocality of the photon itself in the sense that the wavefunction constrained by the RC (\ref{RQC}) does not mean the probability amplitude for the position of the photon.

\section{Nonlocality of photon determined by RC}\label{NP}

\subsection{No probability density exists for position of photon}

Wave equation (\ref{DLE}) can be rewritten as
\begin{equation}\label{DLE-4D}
    \Gamma_0 \frac{\partial \Psi}{\partial x_0}
   +\Gamma_i \frac{\partial \Psi}{\partial x_i}=0.
\end{equation}
Its Hermitian conjugate reads
\begin{equation*}
    -\frac{\partial \Psi^\dag}{\partial x_0} \Gamma_0
    +\frac{\partial \Psi^\dag}{\partial x_i} \Gamma_i =0.
\end{equation*}
Multiplying this equation by $\Gamma_0$ on the right, one has
\begin{equation}\label{DLE-TC}
    \frac{\partial \bar{\Psi}}{\partial x_0} \Gamma_0
   +\frac{\partial \bar{\Psi}}{\partial x_i} \Gamma_i =0,
\end{equation}
where
$\bar{\Psi}=\Psi^\dag \Gamma_0$. Multiplying Eq. (\ref{DLE-4D}) by $\bar{\Psi}$ on the left and Eq. (\ref{DLE-TC}) by $\Psi$ on the right, and summing, one gets the following continuity equation,
\begin{equation*}
    \frac{\partial}{\partial x_\mu} (\bar{\Psi} \Gamma_\mu \Psi)=0.
\end{equation*}
This means that the bilinear quantities
\begin{equation*}
    j_\mu =i c \bar{\Psi} \Gamma_\mu \Psi =i c \Psi^\dag \Gamma_0 \Gamma_\mu \Psi
\end{equation*}
form a four-vector. The time component
$j_0=ic \Psi^\dag \Psi$,
which corresponds to the positive-definite entity
$j_0/ic=\Psi^\dag \Psi$,
thus defines a constant of motion~\cite{Cori-S},
\begin{equation}\label{P}
    P=\int \Psi^\dag \Psi d^3 x .
\end{equation}
This constant of motion can be reasonably interpreted as the total probability of the photon.

But on the other hand, it is seen from the RC (\ref{CE-k}) in momentum representation that
$ \mathbf{f}_u^\ast \cdot \mathbf{f}_u = \mathbf{f}_l^\ast \cdot \mathbf{f}_l $.
Considering this relation, one substitutes Eq. (\ref{FT}) into Eq. (\ref{P}) to get
\begin{equation*}
    P=\int \mathbf{f}_u^\ast \cdot \mathbf{f}_u d^3 k
     =\int \mathbf{f}_l^\ast \cdot \mathbf{f}_l d^3 k.
\end{equation*}
With the help of Eq. (\ref{FT-f}), it is transformed back into
\begin{equation*}
    P=\int \mathbf{F}_u^\ast \cdot \mathbf{F}_u d^3 x
     =\int \mathbf{F}_l^\ast \cdot \mathbf{F}_l d^3 x
\end{equation*}
in position representation, which is expressed in terms of the wavefunction $\Psi$ as
\begin{equation*}
   P=\int \Psi^\dag (1+\Gamma_0) \Psi d^3 x
    =\int \Psi^\dag (1-\Gamma_0) \Psi d^3 x.
\end{equation*}
Since neither $\Psi^\dag (1+\Gamma_0)\Psi$ nor $\Psi^\dag (1-\Gamma_0)\Psi$
is equal to $\Psi^\dag \Psi$ for any particular photon state $\Psi$, no unique probability density exists for the position of the photon.
By this it is meant that the integrand $\Psi^\dag \Psi$ in expression (\ref{P}) cannot be interpreted as the probability density for the position of the photon.

In a word, the RC (\ref{RQC}) renders the photon nonlocal in such a way that the position-representation wavefunction does not mean the probability amplitude for the position of the photon though the integral of its modulus squared over the whole position space gives the total probability.
To the best of my knowledge, this is the first time to connect the nonlocality of the photon with a constraint equation on the wavefunction.
To further appreciate the nonlocality of the photon in quantum mechanics, it is helpful to look at the nonlocal dependence of the upper and lower parts of the quantum wavefunction on the classical electric and magnetic fields.

\subsection{Nonlocal dependence of quantum wavefunction on classical fields}

The real-valued electric and magnetic fields of a free radiation field, when expressed as~\cite{Akhi-B}
\begin{equation}\label{EH-R}
    \boldsymbol{\mathcal E}=\frac{1}{\sqrt 2}(\mathbf{E} +\mathbf{E}^*), \quad
    \boldsymbol{\mathcal H}=\frac{1}{\sqrt 2}(\mathbf{H} +\mathbf{H}^*),
\end{equation}
can be expanded in terms of the plane-wave modes as
\begin{subequations}\label{EH}
\begin{align}
  \mathbf{E} (\mathbf{x},t) &= \frac{1}{(2 \pi)^{3/2}}
 \int\mathbf{e}(\mathbf{k},t)\exp(i\mathbf{k} \cdot \mathbf{x}) d^3 k,\label{CE} \\
  \mathbf{H} (\mathbf{x},t) &= \frac{1}{(2 \pi)^{3/2}}
 \int \mathbf{h}(\mathbf{k},t) \exp(i \mathbf{k} \cdot \mathbf{x}) d^3 k,
\end{align}
\end{subequations}
where the expansion coefficients $\mathbf e$ and $\mathbf h$ are related to each other via
\begin{equation*}
    \mathbf{h}= \frac{1}{\mu_0 c} \mathbf{w} \times \mathbf{e}, \quad
    \mathbf{e}=-\frac{1}{\varepsilon_0 c} \mathbf{w} \times \mathbf{h}
\end{equation*}
by virtue of Maxwell's equations (\ref{ME1})-(\ref{ME2}).
In terms of the expansion coefficients, the spin and orbital angular momentum identified in classical theory are given by~\cite{Li09, Li16E}
\begin{eqnarray*}
  -i \int \frac{\varepsilon_0}{ck} \mathbf{e}^* \times \mathbf{e} d^3 k
   \quad & \mathrm{or} \quad &
     -i \int \frac{\mu_0}{ck}         \mathbf{h}^* \times \mathbf{h} d^3 k, \\
  -i \int \frac{\varepsilon_0}{ck} \mathbf{e}^*
             (\mathbf{k} \times \nabla_{\mathbf k})       \mathbf{e} d^3 k
   \quad & \mathrm{or} \quad &
  -i \int \frac{\mu_0}{ck}         \mathbf{h}^*
             (\mathbf{k} \times \nabla_{\mathbf k})       \mathbf{h} d^3 k,
\end{eqnarray*}
respectively.
If they are postulated to be equal to the expectation values of their counterparts in quantum mechanics, a comparison with Eqs. (\ref{EoS-f}) and (\ref{EoL-f}) leads to
\begin{equation}\label{CR}
    \mathbf{f}_u =\Big( \frac{\varepsilon_0}{\hbar c k} \Big)^{1/2}
                  \mathbf{e},                                          \quad
    \mathbf{f}_l =\Big( \frac{\mu_0}{\hbar c k} \Big)^{1/2} \mathbf{h},
\end{equation}
in consistency with correspondences (\ref{corr}). Since $\frac{1}{\sqrt k}$ is the Fourier transformation of
$\frac{1}{2 |\mathbf{x}|^{5/2}}$,
\begin{equation*}
    \frac{1}{\sqrt k}
   =\frac{1}{(2 \pi)^{3/2}}
    \int \frac{1}{2 |\mathbf{x}|^{5/2}} \exp(-i \mathbf{k} \cdot \mathbf{x}) d^3 x,
\end{equation*}
it follows from Eqs. (\ref{CR}), (\ref{FT}), and (\ref{EH}) that the upper and lower parts of the wavefunction are expressed in terms of the complex vector functions $\mathbf E$ and $\mathbf H$ as~\cite{Cook}
\begin{subequations}\label{Psi}
\begin{align}
   \mathbf{F}_u (\mathbf{x},t) &
  =\sqrt{\frac{\varepsilon_0}{2 \pi \hbar c}} \int
   \frac{\mathbf{E}(\mathbf{x}',t)}{4 \pi |\mathbf{x}-\mathbf{x}'|^{5/2}} d^3 x',\\
   \mathbf{F}_l (\mathbf{x},t) &
  =\sqrt{\frac{\mu_0}{2 \pi \hbar c}} \int \frac{\mathbf{H}(\mathbf{x}',t)}
   {4 \pi |\mathbf{x}-\mathbf{x}'|^{5/2}} d^3 x',
\end{align}
\end{subequations}
respectively. The upper (or lower) part of the quantum wavefunction taken at one particular point $\mathbf x$ depends not only on the value of the classical function $\mathbf E$ (or $\mathbf H$) at that point but also on the value at all other points.
That is to say, the quantum wavefunction does not locally depend on the classical functions $\mathbf E$ and $\mathbf H$.

Furthermore, by denoting by $ \boldsymbol{\varepsilon} (\mathbf{k},t) $ and $\boldsymbol{\eta} (\mathbf{k},t)$ the Fourier coefficients of $\boldsymbol{\mathcal E}$ and $\boldsymbol{\mathcal H}$, respectively,
\begin{subequations}
\begin{align}
  \boldsymbol{\mathcal E} (\mathbf{x},t) &= \frac{1}{(2 \pi)^{3/2}} \int \boldsymbol{\varepsilon} (\mathbf{k},t) \exp(i \mathbf{k} \cdot \mathbf{x}) d^3 k, \label{RE} \\
  \boldsymbol{\mathcal H} (\mathbf{x},t) &= \frac{1}{(2 \pi)^{3/2}} \int \boldsymbol{\eta}    (\mathbf{k},t) \exp(i \mathbf{k} \cdot \mathbf{x}) d^3 k, \label{RH}
\end{align}
\end{subequations}
a comparison with Eq. (\ref{EH-R}) gives
\begin{equation*}
    \boldsymbol{\varepsilon}(\mathbf{k},t) =\frac{1}{\sqrt 2}
    [\mathbf{e}(\mathbf{k},t)+\mathbf{e}^\ast (-\mathbf{k},t)], \quad
    \boldsymbol{\eta}       (\mathbf{k},t) =\frac{1}{\sqrt 2}
    [\mathbf{h}(\mathbf{k},t)+\mathbf{h}^\ast (-\mathbf{k},t)],
\end{equation*}
which have the properties,
\begin{equation}\label{symm}
    \boldsymbol{\varepsilon}^\ast (-\mathbf{k},t)=\boldsymbol{\varepsilon}(\mathbf{k},t), \quad
    \boldsymbol{\eta}^\ast        (-\mathbf{k},t)=\boldsymbol{\eta}       (\mathbf{k},t).
\end{equation}
It is seen that $\mathbf e$ and $\mathbf h$, the Fourier coefficients of the complex functions $\mathbf E$ and $\mathbf H$, are different from $\boldsymbol{\varepsilon}$ and $\boldsymbol{\eta}$. They are not constrained by such conditions as Eqs. (\ref{symm}).
But they can be expressed in terms of $\boldsymbol{\varepsilon}$ and $\boldsymbol{\eta}$ as~\cite{Cohen-DG}
\begin{subequations}\label{RoFC}
\begin{align}
  \mathbf{e}(\mathbf{k},t) & =\frac{1}{\sqrt 2}
    [\boldsymbol{\varepsilon}(\mathbf{k},t)-\frac{\mu_0 c}{k}
     \mathbf{k} \times \boldsymbol{\eta}       (\mathbf{k},t)],        \label{e-epsi+eta}\\
  \mathbf{h}(\mathbf{k},t) & =\frac{1}{\sqrt 2}
    [\boldsymbol{\eta}       (\mathbf{k},t)+\frac{\varepsilon_0 c}{k}
     \mathbf{k} \times \boldsymbol{\varepsilon}(\mathbf{k},t)], \label{h-eta+epsi}
\end{align}
\end{subequations}
by virtue of Maxwell's equations (\ref{ME1})-(\ref{ME2}). According to Eq. (\ref{RH}) and the first equation in (\ref{ME2}), one has
\begin{equation*}
    \frac{1}{(2 \pi)^{3/2}} \int (\mathbf{k} \times \boldsymbol{\eta})
    \exp(i \mathbf{k} \cdot \mathbf{x}) d^3 k
   =-i \varepsilon_0 \frac{\partial \boldsymbol{\mathcal E}}{\partial t}.
\end{equation*}
Since $\frac{1}{k}$ is the Fourier transformation of
$\sqrt{\frac{2}{\pi}} \frac{1}{|\mathbf x|^2}$,
\begin{equation}\label{FI-1/k}
    \frac{1}{k}=\frac{1}{(2 \pi)^{3/2}}
    \int \sqrt{\frac{2}{\pi}} \frac{1}{|\mathbf x|^2}
    \exp(-i \mathbf{k} \cdot \mathbf{x}) d^3 x,
\end{equation}
it follows from Eqs. (\ref{e-epsi+eta}), (\ref{CE}), and (\ref{RE}) that the complex function $\mathbf E$ is related to the electric field $\boldsymbol{\mathcal E}$ in the following way,
\begin{equation}\label{CE-RE}
    \mathbf{E}(\mathbf{x},t)=\frac{1}{\sqrt 2}
   \Big[\boldsymbol{\mathcal E}(\mathbf{x},t)
        +\frac{i}{2 \pi^2 c} \frac{\partial}{\partial t}
         \int \frac{\boldsymbol{\mathcal E}(\mathbf{x}',t)}{|\mathbf{x}-\mathbf{x}'|^2} d^3 x'
   \Big].
\end{equation}
The real part is proportional to the electric field $\boldsymbol{\mathcal E}$ as the first equation in (\ref{EH-R}) requires. But the imaginary part does not locally depend on the electric field.
Similarly, the complex function $\mathbf H$ is related to the magnetic field $\boldsymbol{\mathcal H}$ as follows,
\begin{equation}\label{CH-RH}
    \mathbf{H}(\mathbf{x},t)=\frac{1}{\sqrt 2}
   \Big[ \boldsymbol{\mathcal H}(\mathbf{x},t)
        +\frac{i}{2 \pi^2 c} \frac{\partial}{\partial t}
         \int \frac{\boldsymbol{\mathcal H}(\mathbf{x}',t)}{|\mathbf{x}-\mathbf{x}'|^2} d^3 x'
   \Big],
\end{equation}
where Eq. (\ref{FI-1/k}) as well as the following relation has been used,
\begin{equation*}
    \frac{1}{(2 \pi)^{3/2}} \int (\mathbf{k} \times \boldsymbol{\varepsilon})
    \exp(i \mathbf{k} \cdot \mathbf{x}) d^3 k
   =i \mu_0 \frac{\partial \boldsymbol{\mathcal H}}{\partial t}.
\end{equation*}
In a word, the complex functions $\mathbf E$ and $\mathbf H$ do not locally depend on the electric and magnetic fields.

It is concluded from Eqs. (\ref{Psi}), (\ref{CE-RE}), and (\ref{CH-RH}) that the upper and lower parts of the wavefunction do not locally depend on the electric and magnetic fields.
Eqs. (\ref{Psi}) show that the wavefunction is similar to the so-called Landau-Peierls wavefunction~\cite{Land-P} except that the complex functions $\mathbf E$ and $\mathbf H$ in Eqs. (\ref{Psi}) do not mean the electric and magnetic fields as Pauli~\cite{Pauli} and Bialynicki-Birula~\cite{Bial96} discussed.
It is noted, as Cook~\cite{Cook} showed, that the nonlocal dependence of the wavefunction on the classical fields does not allow to construct a tensor out of the elements of the wavefunction in the way in which the field-strength tensor is constructed out of the elements of the classical fields.
But any way, the nonlocality of the photon in position space does not mean the absence of the photon wavefunction in position representation as was claimed in the literature~\cite{Akhi-B}.

\section{Conclusions and Remarks}

To conclude, I reinterpreted Darwin's equations (\ref{REoM}) in such a way as to meet the need of the quantum mechanics of the photon. I showed for the first time that the photon wavefunction transforms under Lorentz transformation as a spinor. The relativistic nature of the photon is expressed through the constraint equation (\ref{RQC}) on the wavefunction. It is the RC that underlies the nonlocality of the photon in position space, making the wavefunction no longer be the probability amplitude for the position of the photon.
From Darwin's equations I derived the spin of the photon. I found that the RC has a decisive impact on the properties of the photon spin. It determines the operator for the photon spin to be
$\mathbf{\Omega} \cdot \mathbf{w} \mathbf{w}$
in momentum representation. This in turn makes it impossible to introduce the notion of local density for the photon spin in position space. But on the other hand, it allows to use the position-representation wavefunction to express the expectation value of the photon spin as an integral of different integrands over the whole position space. The nonlocality of the spin of the photon is a reflection of the nonlocality of the photon itself.

Due to the RC, the operator for the photon spin in momentum representation does not satisfy the canonical commutation relation.
As a result, the operator for the photon orbital angular momentum in momentum representation does not satisfy the canonical commutation relation, either, if the total angular momentum is to satisfy the canonical commutation relation~\cite{van-N1, van-N2}.
The only explanation for this is that the RC makes the photon position, represented by $\mathbf{x}=i \nabla_k$ in momentum representation, not satisfy
$[x_i, x_j]=0$.
Otherwise, this equation together with the commutation relations (\ref{CR-PandX}) would lead to the canonical commutation relation of the orbital angular momentum~\cite{Saku}, $[L_i, L_j]=i \hbar \epsilon_{ijk} L_k$.
The conclusion that the wavefunction is not the probability amplitude for the position of the photon is compatible with the non-commutativity of the photon position.
Further discussions are beyond the scope of present paper.

\section*{Acknowledgments}

The author is indebted to Kang-Rui Liu for his helpful discussions. This work was supported in part by the program of Shanghai Municipal Science and Technology Commission under Grant 18ZR1415500.

\appendix

\section{Lorentz Transformation for the Wavefunction}\label{LT}

For simplicity I consider a Lorentz transformation of velocity $v$ along the first axis,
\begin{equation}\label{SLT}
    x'_\mu = a_{\mu \nu} x_\nu,
\end{equation}
for which the transformation matrix assumes the form
\begin{equation*}
    (a_{\mu \nu})=\left(
                    \begin{array}{cccc}
                      \gamma         & -i \beta \gamma & 0 & 0 \\
                      i \beta \gamma & \gamma          & 0 & 0 \\
                      0              & 0               & 1 & 0 \\
                      0              & 0               & 0 & 1 \\
                    \end{array}
                  \right),
\end{equation*}
where $\beta=v/c$ and $\gamma=1/(1-\beta^2)^{1/2}$. I will show that under this transformation, Darwin's equations (\ref{REoM}) are invariant if the wavefunction transforms as follows,
\begin{equation}\label{LT-WF}
    \Psi' (x'_\mu) =\Lambda \Psi (x_\mu),
\end{equation}
where
$\Lambda=\exp(-i  \chi \Gamma_0 \Gamma_1)$
and $\chi= \cosh^{-1} \gamma$.

To this end, it is helpful to write Eq. (\ref{LT-WF}) out explicitly in terms of the upper and lower parts of $\Psi'$ and $\Psi$.
Letting
$\Psi'=\frac{1}{\sqrt 2} \bigg(
                           \begin{array}{c}
                             \mathbf{F}'_u \\
                             \mathbf{F}'_v \\
                           \end{array}
                         \bigg)
$ and noting that
\begin{equation*}
    \Lambda=1-i \Gamma_0 \Gamma_1 \sinh \chi -\Gamma_1^2 (1-\cosh \chi),
\end{equation*}
one can rewrite Eq. (\ref{LT-WF}) as
\begin{equation*}
    \bigg(
      \begin{array}{c}
        \mathbf{F}'_u \\
        \mathbf{F}'_v \\
      \end{array}
    \bigg)
   =\bigg(
      \begin{array}{cc}
        1-\Sigma_1^2 (1-\cosh \chi) & -i \Sigma_1 \sinh \chi \\
        i \Sigma_1 \sinh \chi & 1-\Sigma_1^2 (1-\cosh \chi)  \\
      \end{array}
    \bigg)
    \bigg(
      \begin{array}{c}
        \mathbf{F}_u \\
        \mathbf{F}_v \\
      \end{array}
    \bigg).
\end{equation*}
The components of the transformed wavefunction are therefore related to those of the original wavefunction via the following formulae,
\begin{subequations}\label{LT-1}
\begin{align}
  F'_{u1} & = F_{u1},     \\
  F'_{u2} & = F_{u2} \cosh \chi -F_{v3} \sinh \chi,     \\
  F'_{u3} & = F_{u3} \cosh \chi +F_{v2} \sinh \chi,
\end{align}
\end{subequations}
and
\begin{subequations}\label{LT-2}
\begin{align}
  F'_{v1} & = F_{v1},     \\
  F'_{v2} & = F_{v2} \cosh \chi +F_{u3} \sinh \chi,     \\
  F'_{v3} & = F_{v3} \cosh \chi -F_{u2} \sinh \chi.
\end{align}
\end{subequations}
Interestingly, they are the same as the transformation formulae~\cite{Jack} for the electric and magnetic fields of a free radiation under Lorentz transformation (\ref{SLT}). The invariance of Darwin's equations (\ref{REoM}) under Lorentz transformations (\ref{SLT})-(\ref{LT-WF}) is shown below.

\section{Lorentz Covariance of Darwin's Equations}\label{LC}

Suppose that one has Darwin's equations in the primed system,
\begin{subequations}
\begin{align}
  \Gamma_\mu p'_\mu \Psi' & =0,                                \label{DLE'} \\
  (\mathbf{\Gamma} \cdot \mathbf{p}')^2 \Psi' & =p'^2 \Psi'.   \label{RQC'}
\end{align}
\end{subequations}
Multiplying the wave equation (\ref{DLE'}) by $\Gamma_0$ on the left, one obtains
\begin{equation*}
    p'_0 \Psi' +(\Gamma_0 \mathbf{\Gamma} \cdot \mathbf{p}') \Psi' =0.
\end{equation*}
Upon substituting Eq. (\ref{LT-WF}), one gets
\begin{equation*}
    p'_0 \Lambda \Psi +(\Gamma_0 \mathbf{\Gamma} \cdot \mathbf{p}') \Lambda \Psi =0,
\end{equation*}
which is equivalent to
\begin{equation*}
    p'_0 \Psi +\Lambda^{-1} (\Gamma_0 \mathbf{\Gamma} \cdot \mathbf{p}') \Lambda \Psi =0.
\end{equation*}
Noticing that
\begin{eqnarray*}
  \Lambda^{-1} \Gamma_0 \Gamma_1 \Lambda &=& \Gamma_0 \Gamma_1, \\
  \Lambda^{-1} \Gamma_0 \Gamma_2 \Lambda &=& \gamma \Gamma_0 \Gamma_2
  -i \beta \gamma (\Gamma_1 \Gamma_2 -\Gamma_2 \Gamma_1), \\
  \Lambda^{-1} \Gamma_0 \Gamma_3 \Lambda &=& \gamma \Gamma_0 \Gamma_3
  -i \beta \gamma (\Gamma_1 \Gamma_3 -\Gamma_3 \Gamma_1),
\end{eqnarray*}
and considering
$p'_\mu =a_{\mu \nu} p_\nu$, one finds after lengthy but straightforward algebra,
\begin{equation}\label{IL}
    [(\Gamma_0-i \beta \Gamma_1)\Gamma_\mu p_\mu
    -i \beta (p_1 -\mathbf{\Gamma} \cdot \mathbf{p} \Gamma_1)] \Psi=0.
\end{equation}
It is noted that the second term of this equation on the left reads
\begin{equation*}
    (p_1 -\mathbf{\Gamma} \cdot \mathbf{p} \Gamma_1) \Psi
   =\bigg(
      \begin{array}{c}
        (p_1-\mathbf{\Sigma \cdot \mathbf{p}} \Sigma_1) \mathbf{F}_u \\
        (p_1-\mathbf{\Sigma \cdot \mathbf{p}} \Sigma_1) \mathbf{F}_v \\
      \end{array}
    \bigg),
\end{equation*}
which reduces to
\begin{equation}\label{CC}
    (p_1 -\mathbf{\Gamma} \cdot \mathbf{p} \Gamma_1) \Psi
   =\bigg(
      \begin{array}{c}
        (\mathbf{p} \cdot \mathbf{F}_u) \mathbf{e}_1 \\
        (\mathbf{p} \cdot \mathbf{F}_v) \mathbf{e}_1 \\
      \end{array}
    \bigg)
\end{equation}
in accordance with the relation
$(\mathbf{\Sigma} \cdot \mathbf{a}) \mathbf{b}= i \mathbf{a} \times \mathbf{b} $.

On the other hand, as is discussed in Section \ref{LCoDE}, the RC (\ref{RQC'}) in the primed system means
\begin{equation*}
    \nabla' \cdot \mathbf{F}'_u=\frac{\partial}{\partial x'_i} F'_{ui}=0, \quad
    \nabla' \cdot \mathbf{F}'_v=\frac{\partial}{\partial x'_i} F'_{vi}=0.
\end{equation*}
By using Lorentz transformation (\ref{SLT}) and transformation formulae (\ref{LT-1})-(\ref{LT-2}), it is straightforward to show that
\begin{equation*}
    \nabla' \cdot \mathbf{F}'_u=\gamma \nabla \cdot \mathbf{F}_u, \quad
    \nabla' \cdot \mathbf{F}'_v=\gamma \nabla \cdot \mathbf{F}_v.
\end{equation*}
As a result, one must have
\begin{equation}\label{ZD}
    \nabla \cdot \mathbf{F}_u=0, \quad \nabla \cdot \mathbf{F}_v=0,
\end{equation}
which means the RC
\begin{equation*}
    (\mathbf{\Gamma} \cdot \mathbf{p})^2 \Psi =p^2 \Psi
\end{equation*}
in the unprimed system. This shows that the RC (\ref{RQC}) is invariant under Lorentz transformations (\ref{SLT})-(\ref{LT-WF}). With the help of Eqs. (\ref{ZD}), Eq. (\ref{CC}) reduces to
$(p_1 -\mathbf{\Gamma} \cdot \mathbf{p} \Gamma_1) \Psi=0$
and hence Eq. (\ref{IL}) becomes
\begin{equation*}
    (\Gamma_0-i \beta \Gamma_1)\Gamma_\mu p_\mu \Psi=0.
\end{equation*}
Because the matrix $\Gamma_0-i \beta \Gamma_1$ is  invertible, one finally gets for the wave equation in the unprimed system,
\begin{equation*}
    \Gamma_\mu p_\mu \Psi=0.
\end{equation*}
In a word, the wave equation (\ref{DLE}) is invariant under Lorentz transformations (\ref{SLT})-(\ref{LT-WF}).


\end{document}